\documentclass[onecolumn,secnumarabic,
amsmath,amssymb,nofootinbib,
 aps, pre]{revtex4-2}

\usepackage[mathlines]{lineno}
\usepackage{graphicx}
\usepackage{bm}
\usepackage[capitalize]{cleveref}
\usepackage{siunitx}
\expandafter\ifx\csname package@font\endcsname\relax\else
 \expandafter\expandafter
 \expandafter\usepackage
 \expandafter\expandafter
 \expandafter{\csname package@font\endcsname}%
\fi
\usepackage{dcolumn}
\usepackage{titlesec}

\graphicspath{{figs/}}

\begin{document}

\title{Fluid flow in 3-dimensional porous systems shows power law scaling with Minkowski functionals} 

\author{R. A. I. Haque$^{1,2}$, A. J. Mitra$^3$ and T. Dutta$^{1,2}$}
\email[Tapati Dutta ]{  tapati$\_$dutta@sxccal.edu}
\affiliation{$^1$Physics Department, St. Xavier's College, Kolkata 700016, India\\
$^2$Condensed Matter Physics Research Centre, Jadavpur University, India\\
$^3$Mathematical Sciences, Montana Tech, Butte, MT 59701, United States\\}
\date{\today}%

\begin{abstract}
Integral geometry uses four geometric invariants -- the Minkowski functionals -- to characterize certain subsets of 3-dimensional space. The question was, how is the fluid flow in a 3-dimensional porous system related to these invariants? In this work, we systematically study the dependency of permeability on the geometrical characteristics of two categories of 3-dimensional porous systems generated: (i) stochastic and (ii) deterministic. For the stochastic systems, we investigated both normal and log-normal size distribution of grains. For the deterministic porous systems, we checked for a cubic and a hexagonal arrangement of grains of equal size. Our studies reveal that for any 3-dimensional porous system, ordered or disordered, permeability $k$ follows a unique scaling relation with the Minkowski functionals: (a) volume of the pore space, (b) integral mean curvature, (c) Euler Characteristic and (d) critical cross-sectional area of the pore space. The cubic and the hexagonal symmetrical systems formed the upper and lower bounds of the scaling relations, respectively. The disordered systems lay between these bounds. Moreover, we propose a combinatoric $F$ that weaves together the four Minkowski functionals and follows a power-law scaling with permeability. The scaling exponent is independent of particle size and distribution and has a universal value of $ 0.428$ for 3-dimensional porous systems built of spherical grains.
\end{abstract}

\keywords{Porous media, Fluid flow, Scaling, Geometric invariants, 3-dimensions, Minkowski functionals}
\maketitle

\tableofcontents

%\linenumbers\relax

%\maketitle
%\newpage

\section{Introduction}
Understanding flow through porous medium continues to be an active research area because of its important applications in our daily life. From subsurface flow important to agriculture, oil, and natural gas harvesting, $CO_2$ sequestration in sedimentary rocks to engineering applications, fluid transport in the pore space of a granular 3-dimensional structure is a complex non-linear problem whose solution largely depends on modeling and simulation. Desktop experiments cannot faithfully reproduce in-situ conditions and can, at best, act as pointers to solutions. Transport of fluids in real situations, e.g., sedimentary rocks, is characterized by permeability or conductivity of the system. However, these macro properties of the porous system are guided by micro to mesoscale properties like the size and shape of grains, which can have a wide distribution. The wetting property of the fluid in the matrix is affected by surface tension, and viscosity determines whether the fluid shall show both capillary and Saffman-Taylor like instabilities \cite{Lenormand1989, Saffman1958}.

The continuum-scale models that relate permeability and capillary pressure are based on the assumption of fluid saturation \cite{Dullien2012, Bear2012} and the results can be both material and process dependent \cite{Leverett1941, Genuchten1980}. In order to develop constitutive models that are applicable to more general situations, a geometric characterization of the pore space and its relation to transport properties is desired. Efforts to link the permeability of a porous system to the topological measure Euler characteristic of the porous system have been made in a 2-dimensional porous medium \cite{Scholz2012} where grain overlap is considered to be the most important factor in determining the permeability of the porous system. However, in actual 3-dimensional systems, overlapping grains are not meaningful. Katz-Thomson \cite{Katz1986} proposed that permeability $k$ and the electrical conductivity $\sigma$ of a porous system is linked via a characteristic length $l_c$ of the system by 
\begin{equation}
k = c{l_{c}}^{2}(\frac{\sigma}{\sigma_{0}})
\label{KT}
\end{equation}
where $c$ is a pore geometry dependent constant and $l_{c}$ is a critical pore diameter that cuts off any particle of average diameter greater than $l_{c}$ from percolating through the system, and $\sigma_{0}$ the fluid conductivity. Archie's law that provides an empirical relation between conductivity and the porosity of a porous rock is limited by its validity near the percolation threshold of the porous medium, has found success in simulation studies \cite{Kirkpatrick1973, Halperin1985, Tobochnik1990}, with some exceptions \cite{Sen1985, Bunde1986, Octavio1988}. However, interfacial energy and fluid topology are important parameters that affect fluid flow in porous systems \cite{Joekar2013, Berg2013, Blunt2013, Rucker2015, Held2001}. Results using integral geometry have been used to find a more general solution to the problem \cite{Vogel2010,  Hilfer2002, Arns2010, Schluter2011, Lehmann2008, Schroder2013, Armstrong2019, Purswani2020, Slotte2020}. Mclure et al. \cite{McClure2020} have used a non-dimensional relationship based on the Minkowski–Steiner formula \cite{Federer1959} to predict fluid flow based on geometric factors of several porous rocks.

Our work in this paper is inspired by Hadwiger's characterization theorem that roughly says that the structure of finite unions of convex subsets of 3-dimensional systems can be described by at most four invariant measures \cite{Hadwiger2013}, the Minkowski functionals. Therefore, it is not unreasonable to expect that all transport properties, e.g., permeability, may be linked to the basic invariant geometric measures of the porous medium, as the grains can be assumed to be roughly convex. For this purpose, we simulated 3-dimensional porous structures, both stochastic and deterministic, following certain construction algorithms. The structures have a distribution of particle sizes about a mean size for the stochastic structure generation; the deterministic structures were generated with particles of constant size whose values varied over a range. We determined the permeability of all generated structures by simulating incompressible fluid flow equations conserving mass and momentum under constant pressure gradients. The invariant topological measure of the pore space was determined by the computation of Euler Characteristic $\chi$. The characteristic critical length $l_c$, the integral mean curvature $H$, and the average surface cross-sectional area of the pore space were computed for every porous structure studied. Our study indicated that the fluid flow characteristic, permeability, exhibited a power-law variation with each of the geometric invariants of the porous systems.  This unilateral behaviour of the invariants with permeability prompted us to build an empirical relationship between the permeability $k$ and a combinatoric function $F$ - built from the four geometrical characteristics of the porous system - that displayed a power-law behaviour irrespective of the pore distribution in the system.  The exponent of this scaling relationship was almost identical for both a 3-dimensional self-organized disordered system at equilibrium and a 3-dimensional  deterministic system that displayed certain geometrical symmetry. To the best of our knowledge, this is the first time that a single relationship combining the fluid flow with all the Minkowski functionals has been proposed with a unique exponent for 3-dimensional disordered porous systems. The relationship was established by exploring piecemeal-wise the variation of permeability with each of the geometric descriptors of the system. The cubic and the hexagonal symmetrical systems formed the upper and lower bounds of the scaling relations, respectively. The disordered systems lay between these bounds. Our proposed relations were robust to variations of size and micro-geometry of the pore space, with an exponent that remained constant for both the stochastic and deterministic systems. 

In the following sections, we shall present the methodology of structure construction for both stochastic and deterministic 3-dimensional porous systems, the determination of the flow characteristic $k$, and the computation of each of the invariant geometrical measures characteristic of every structure. This will be followed by the results and the discussion leading to the establishment of the scaling relationship. Finally we shall summarize our findings and future plans as conclusions of this work. 

\section{Sample generation}
We generate a porous stochastic structure in 3-D and simulate the flow of a single fluid through it using a numerical finite difference solution of the steady state Navier Stokes' equation. Most naturally occurring porous rocks have a log-normal particle size distribution \cite{DEXTER1972}. Thus, in order to generate a 3-dimensional porous matrix, spherical particles of different radii  $R$ were chosen randomly from a log-normal distribution with mean radius $\langle R \rangle$ and standard deviation $s$.  The radii of the particles lay within  $\langle R \rangle/4$ to $5\langle R \rangle$ in order to avoid too big or too small particles in the system. Another class of 3-dimensional disordered systems was built for a normal distribution of particle sizes.  The particles were allowed to settle under gravity, filling an imaginary cubical box of size $L = 0.5$ cm. Distinct Element Method (DEM) was used to calculate the forces acting on the particles during the structure generation. It is assumed that:
\begin{itemize}
\item The particles can both translate and rotate independently of each other.
\item Two particles $A$ and $B$ are said to be in contact if the distance $d$ between their centres  satisfies the condition $d \leq R_{A} + R_{B}$.
\item Particles interact via contact points only, where a contact comprises of only two particles. Two particles in contact define a point of intersection in the matrix.
\item Particles are allowed to overlap over a small region at the point of contact; however, the overlaps are very small in relation to particle size. 
\item Newton's second law is used to determine the translational and rotational motion of each particle, while the force-displacement law is used to update the contact forces arising from the relative motion at each contact.
\item Dynamics is implemented by updating particle position in a time step $\Delta t$ small enough to assume constant velocity and acceleration values.
\item $\Delta t$ is chosen in a manner such that disturbances due to a particle cannot propagate further than its nearest neighbours.
\end{itemize}
The particles are allowed to fall freely under gravity along the z-direction from any position $r_{i,j}$ that is chosen stochastically; here, $i$ and $j$ refer to $x$ and $y$ coordinate $\leq L$.
The interaction between the particles is treated as a dynamic process with equilibrium states occurring whenever internal forces become zero. When one particle hits any other particle of the structure, it causes a disturbance. The particles move relative to each other, with the speed of propagation being dependent on particle contact distribution and material properties. The calculation of the net normal force and the net shear force acting on each particle follows the scheme described by Potyondy et al. \cite{Potyondy2004}.

Different samples can be generated by changing the $\langle R \rangle$ and the standard deviation $s$. Depending on  $\langle R \rangle$, the number $N$ of particles that generate the sample is $\approx 2000$.

The deterministic 3-dimensional structures were generated with spheres of equal radius but contained in the imaginary cubical boxes of size $L$. We examined the deterministic structures for two symmetric arrangements: (i) hexagonal and (ii) cubic. For every symmetry, $6$ different radii were chosen for sample generation. 

 \section{Geometric characterization of porous medium}
 
Regardless of the total porosity of the samples, we considered the \textbf{effective porosity} $\phi_{0}$ of the sample spanning channels,  as this porosity alone contributes to fluid flow. To identify sample spanning clusters and the geometrical characteristics of the system, we superposed a $256 \times 256 \times 256$ cubic grid on the 3-dimensional structure with grid length $\delta x \sim 0.002 $ cm. The porous structure was discretized by assigning a grid cell a value of \(1\) if at least $50\%$ of the cell was filled by the matrix. Otherwise, the grid cell was given a value of \(0\). This process converted the porous structure into a binary format.

An effective \textbf{critical area} $A_{c} = \frac{\pi}{4}{l_{c}}^{2}$ was calculated after the determination of the critical length scale $l_c$ - defined as the maximum diameter of a spherical particle that can percolate through the system spanning channels. After all the system-spanning channels were identified via the Hoshen-Koppelmann algorithm, a sorting algorithm was used to determine $l_c$ for each sample.

The topological invariant, \textbf{Euler Characteristic $\chi$} defined as an alternating sum of Betti numbers :  
\begin{equation}
\chi = \beta_{0} - \beta_{1} + \beta_{2} - \cdots
\label{euler}
\end{equation}
where the $\beta_{0}$ represents the number of connected components, $\beta_{1}$ is the 1-dimensional holes or loops,  $\beta_{2}$ represents the 2-dimensional voids or cavities, and so on. 
For 3-dimensional porous systems constructed only by spheres Eq.(\ref{euler}) can be simplified to \cite{Vogel2002}
\begin{equation}
\chi = M - I + N
\label{simple-euler}
\end{equation}
where $M$ is the number of isolated pores, $I$ refers to the number of intersections between grains, i.e., the number of points where two grains touch each other at a point, and $N$ is the number of grains completely enclosed by the pores. The Euler characteristic can be considered to be a measure of connectivity that yields positive values for structures with low connectivity, where \( M \) (isolated pores)  exceeds \( I \) (intersections), and negative values for more highly connected structures, where \( M \) is less than \( I \). In the DEM scenario of particles falling under gravity for the 3-dimensional construction, there is a finite probability of the intersections being circles. However, the particle overlap is assumed to be negligibly small compared to particle size and hence neglected. The Euler Characteristic $\chi$ provides a measure of the connectivity in the sample and is very relevant for fluid flow studies.

The integral \textbf{mean curvature $H$ } of a particle is given by the surface integral
\begin{equation}
H = \int\frac{\kappa_{1} + \kappa_{2}}{2} ds
\label{curv1}
\end{equation} 
where $\kappa_{1}$ and $\kappa_{2}$ are the principal radii of curvature of the grain. For a spherical grain $\kappa_{1} = \kappa_{2} = \frac{1}{R} $. The mean curvature of the void surface of an assembly of spherical particles is
\begin{equation}
H = \sum_{i}\frac{1}{R_i}4\pi{R_{i}}^2\\
  = 4\pi\sum_{i}R_{i}
\label{curv2}
\end{equation}

\section{Fluid transport}
Fluid transport in a porous structure under a suitable pressure gradient is described by the Navier-Stokes equation 
\begin{equation}
\rho\frac{\delta \textbf{V}}{\delta t} + (\textbf{V}.\nabla)\textbf{V} + \nabla P - \mu \nabla^2\textbf{V} = f_e
\label{ns} 
\end{equation}
where $\textbf{V}$, $P$ and $f_e$ represent the velocity, pressure, and external force per unit volume respectively, $\rho$ and $\mu$ are respectively the density
and dynamic viscosity of the fluid. Neglecting the inertial term and assuming no external forces acting on the fluid, eq.(\ref{ns})
simplifies to
\begin{equation}
\frac{\delta \textbf{V}}{\delta t}= -\frac{1}{\rho}\nabla P + \eta \nabla^2\textbf{V}
\label{ns1}
\end{equation}
where $\eta = \frac{\mu}{\rho}$ is the kinematic viscosity. For an incompressible fluid, the equation of continuity is
\begin{equation}
\nabla.\textbf{V}=0
\label{cont}
\end{equation}
Eqs.(\ref{ns1}) and (\ref{cont}), when solved together, give the steady state condition of flow in the structure.

The Hoshen and Kopelman algorithm \cite{Hoshen1976} was used to identify the channels spanning the sample. 
The pressure and velocity fields were solved by the procedure described by Sarkar et al.  \cite{Sarkar2004} with some necessary departures
appropriate to our problem.  
The space and time discretized versions of Eqs.(\ref{ns1}) and (\ref{cont}) were used iteratively to obtain the steady-state flow \cite{Sadhukhan2007}. This was identified when the difference of velocity between successive time steps of iteration was $10^{-9}$ or less.
Our simulation was on a one time injection of fluid. The steady-state values of velocity and pressure at all points of the spanning transport channels were noted.
The permeability was calculated according to Darcy's law
\begin{equation}
\textbf{q} = \frac{k}{\mu} \nabla P
\label{darcy} 
\end{equation} 
where $\textbf{q}$ is the flux, $k$ the permeability, $\mu$ the viscosity and $\nabla P$ the pressure gradient across the sample.

\section{Results and Discussion}
\begin{figure}[h!]
\includegraphics[width=0.65\textwidth]{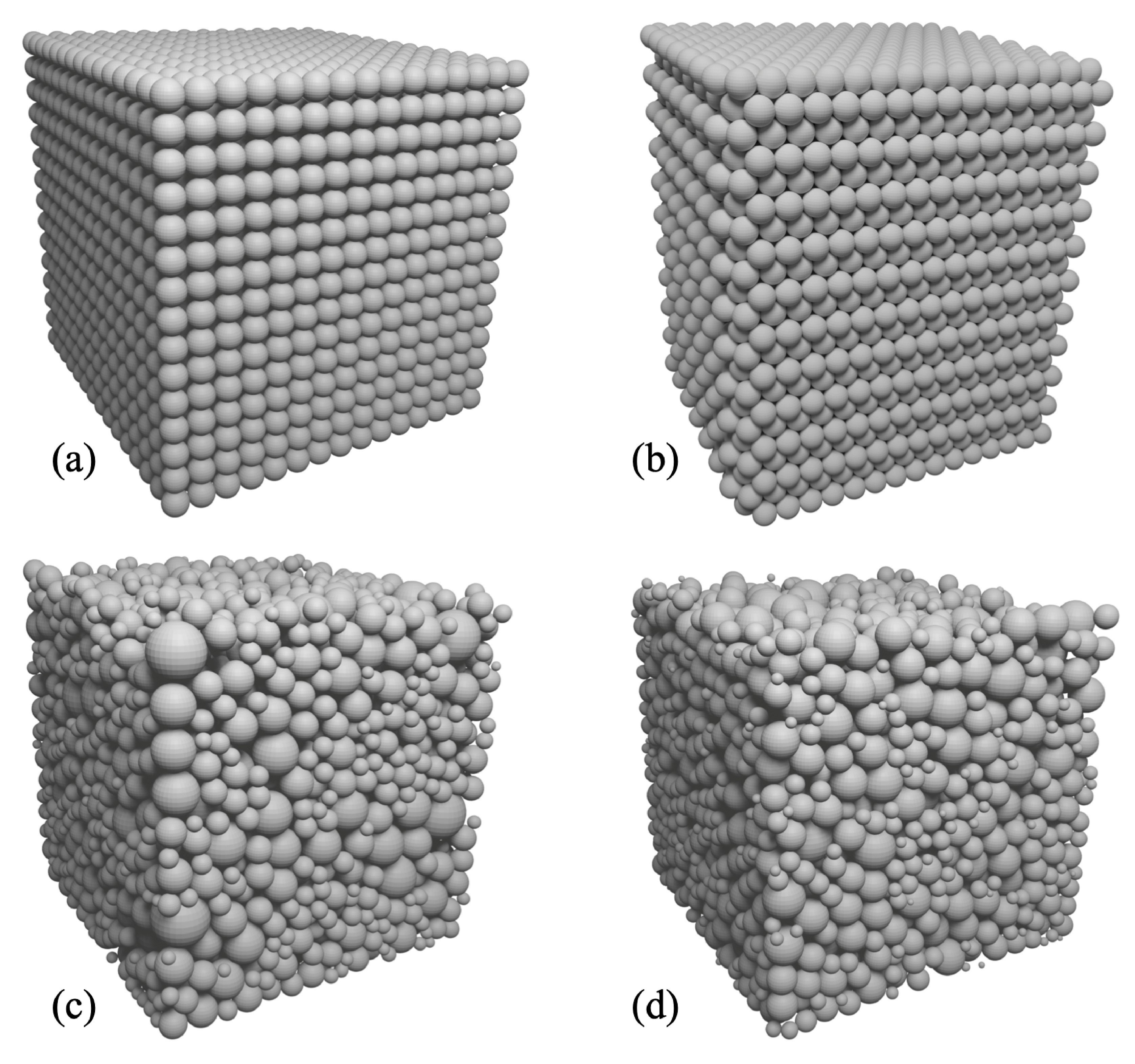}
\caption{3-dimensional porous structures generated using DEM. (a) Deterministic structure having cubic symmetry (b) Deterministic structure having hexagonal symmetry (c) Stochastic structure with particle size chosen from a log-normal distribution. (d) Stochastic structure with particle size chosen from a normal distribution.}
\label{structure}
\end{figure}

 The 3-dimensional porous structures were constructed as discussed earlier. We worked with two stochastic structures with particle size distributions chosen from (i) log-normal and (ii) normal distributions. The distributions were built around mean particle sizes $\langle R \rangle$ varying from \SI{0.02}{\cm} to \SI{0.045}{\cm} and $s$ varying from \SI{0.0025}{\cm} to \SI{0.015}{\cm}, and all results averaged over $30$ configurations. As the length $L$ of the macroscopic cubic structure was kept constant, the number $N$ of particles varied from $419$ to $3656$ depending on the particle sizes. We also worked on two deterministic porous structures having (i) cubic and (ii) hexagonal symmetries. Figure.(\ref{structure}) displays typical images of the four different types of 3-dimensional porous structures generated for this study.

\begin{figure}[h!]
\includegraphics[width=0.75\textwidth]{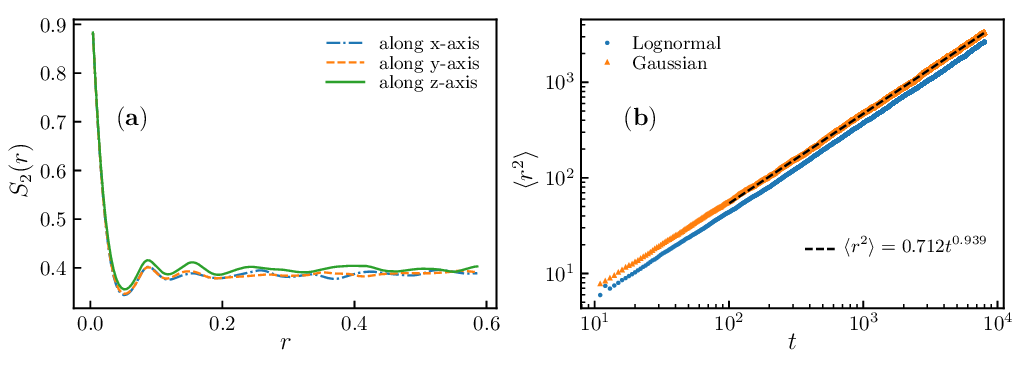}
\caption{Typical micro-geometric characteristic of 3-dimensional stochastic porous systems. Lognormal stochasticity is shown here. (a) Variation of 2-point correlation function $S_{2}(r)$ with $r$ along principle axes. (b) Mean square displacement $\langle r^{2} \rangle$ with time $t$ shows subdiffusive behaviour.}
\label{s2_diff_lc}
\end{figure}

 After structure generation, the sample spanning void clusters were identified to study fluid transport. The direction of the pressure gradient was identified as the z-axis of the system and was identical to the direction of particle deposition.  For the stochastic structures, we determined the 2-point density correlation function $S_{2}(r)$, defined by
 \begin{equation}
 S_{2}(r) =  \langle p(r^{\prime}) p(r + r^{\prime})\rangle
 \label{s2}
 \end{equation}
where $p(r)$ defines the probability of finding a void at position $r$. For the stochastic systems, the variation of $S_{2}(r)$ versus $r$ computed along each axes for a typical sample, Fig.(\ref{s2_diff_lc}a), indicates that the sample was isotropic in the transverse (x-y) plane. A slight anisotropy along the z-axis, the direction of grain deposition, is indicated. 

 To understand the pore micro-geometry, we studied diffusion using a random walk algorithm; details are available in \cite{Giri2012}. The mean square displacement $\langle r^{2} \rangle$  showed a power law variation with time $t$, Fig.(\ref{s2_diff_lc}b), indicative of subdiffusive behaviour.

The geometric characteristics of the Euler Characteristic $\chi$, the mean integral curvature, and the effective porosity were determined for every sample generated, as discussed in the earlier section. Finally, the fluid flow measure permeability ($k$) of each sample was determined. The space and time discretized versions of Eqs.(\ref{ns1}) and (\ref{cont}) were used iteratively to obtain the steady-state flow when the pressure and velocity values at every point in the sample spanning void cluster were known.
\begin{figure}[h!]
\includegraphics[width=0.75\textwidth]{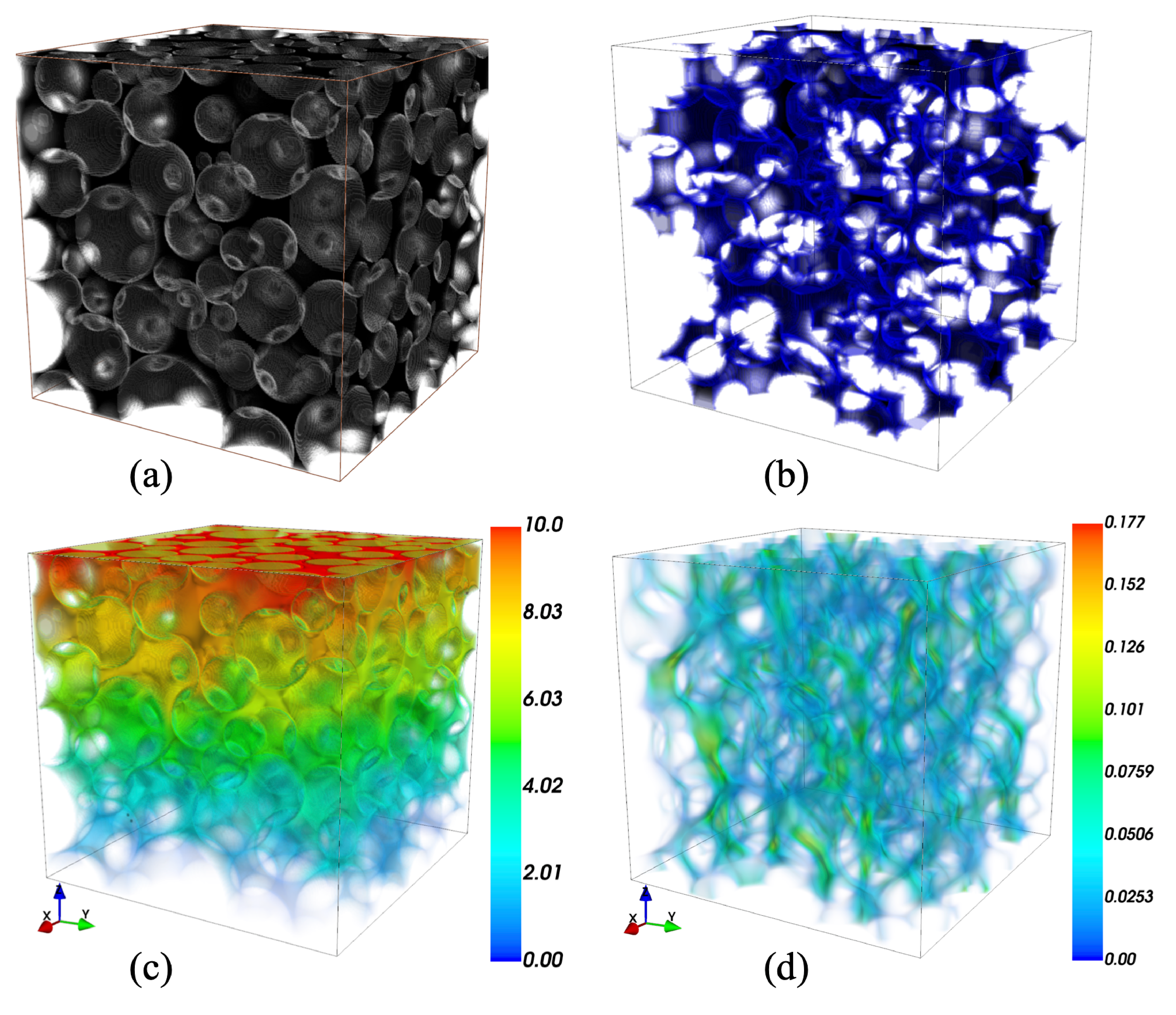}
\caption{(a) Porous structure of a log-normal stochastic sample, (b) the system spanning path through which a sphere of diameter $l_{c}$ can freely move, (c) pressure distribution inside the porous space,  (d) steady-state velocity profile across the sample. The colour legends provide the value scales in CGS units.}
\label{press-vel}
\end{figure}
 
To have an idea of the pore space of our generated systems, we have shown the porous structure of a disordered sample having log-normal particle size distribution, Figure(\ref{press-vel}a). Figure (\ref{press-vel}b) shows the percolating path through which a sphere of diameter $l_{c}$ can freely move.
 Figures.(\ref{press-vel}c and d) show the pressure profile and the velocity field at the steady state when a fluid is allowed to flow under a constant pressure gradient. The output fluid flux was determined, and the permeability of each sample was computed using Eq.(\ref{darcy}).

\begin{figure}[h!]
\includegraphics[width=0.75\textwidth]{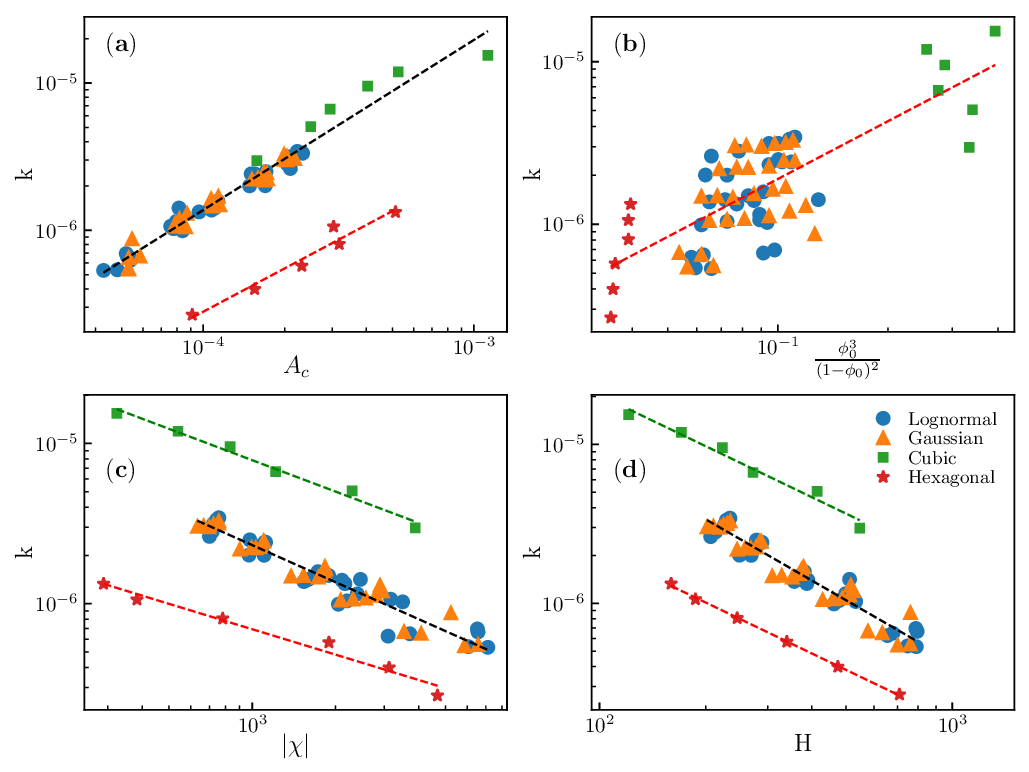}
\caption{(a) The log-log variation of permeability with $A_{c}$ for the deterministic and stochastic 3-dimensional porous structures, (b) variation of permeability with porosity $\phi_{0}$, (c) the log-log variation of permeability with the Euler Characteristic $\chi$,  (d) the variation of permeability with the mean integral curvature.}
\label{k-l_cH}
\end{figure} 
Hadwiger’s characterization theorem states that, at the most, only four invariant measures -- (i) area, (ii) volume, (iii) the Euler Characteristic, and (iv) integral mean curvature -- are required to characterize a 3-dimensional system formed by a union of convex solids. Our intuition suggested that permeability could be expressed as a function of the Minkowski functionals: (i) square of the characteristic length $l_{c}$ (ii) the integral mean curvature $H$, (iii) the topological measure of Euler Characteristic $\chi$, and (iv) the effective porosity $\phi_{0}$ which is defined as the volume of the porous channel scaled by the system volume which is a constant for all the cases considered. To this end, we examined the variation of the permeability with each of these measures. The variation of permeability showed a power law dependence with $l_{c}^{2}$. It may be noted that the effective cross-section of the transport channel $ A_{c} = \frac{\pi}{4} l_{c}^{2}$. Thus the variation of $k$ with $A_{c}$ shows a scaling behaviour of the form, Fig.(\ref{k-l_cH}a) 
\begin{equation}
k = C_{1} {A_{c}}^{m_{1}}
\label{k-lc}
\end{equation}
with $m_{1} \approx 1.066$. Permeability $k$ increases with increasing $A_{c}$, i.e., larger pore throats aided permeability as expected. 

When the variation of permeability with effective porosity was checked for the system, Fig.(\ref{k-l_cH}b), it was observed that the data points formed a cloud. However, the clouds for the different systems showed a power-law scaling with permeability with an exponent $m_{2} \approx 1.180$.
\begin{equation}
k = C_{2} \left(\frac{\phi_{0}^{3}}{(1 - \phi_{0})^{2}}\right)^{m2}
\label{k-phi0}
\end{equation}

\begin{figure}[h!]
\includegraphics[width=0.75\textwidth]{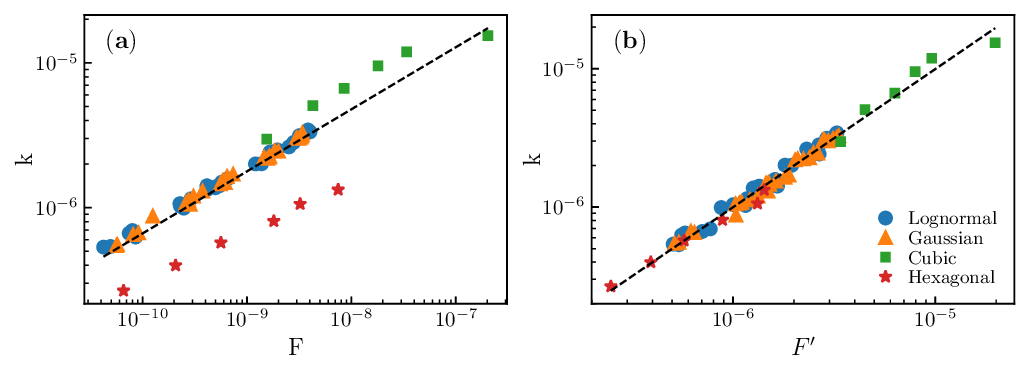}
\caption{(a) Variation of permeability $k$ with $F$, a build-up function of the geometric invariants of the samples. (b)Variation of permeability $k$ with $F$ using regression shows a very good collapse on a single power-law scaling function irrespective of sample type.}
\label{k-chi2_collapse}
\end{figure} 

When the behaviour of permeability was checked for variation in the topological connectivity measure the Euler Characteristic $\chi$, permeability decreased with $| \chi |$ following a scaling behaviour of the the form, Fig.(\ref{k-l_cH}c): 
\begin{equation}
k = C_{3} |\chi|^{- m_{3}}
\label{k-EC}
\end{equation}
 The exponent $m_{3} \approx 0.648$.
 
 Variation of permeability with the integral mean curvature $H$ followed a scaling law of the form
\begin{equation}
k = C_{4} H^{- m_{4}}
\label{k-H}
\end{equation}
The exponent $m_{4}$ was almost identical for both the deterministic and the stochastic systems, being $m_{4} \approx 1.17$, Figs.(\ref{k-l_cH}d). Thus, smaller curvatures, i.e., larger spherical grains, left bigger voids in the system that were conducive to fluid flow.
 
The common features of our investigations so far have been:\\
\begin{itemize}
\item Permeability showed a power law dependence with each of the different geometric measures of the porous system, having the form:
\begin{equation}
k = Cx^{m}
\end{equation}
with $C$ a constant,  $x$ being any one of the geometric measures of the 3-dimensional system, and $m$ being the corresponding exponent.

\item In all the relations investigated, Eqs.(\ref{k-lc} to \ref{k-EC}), the exponents had, on average, identical values for all the systems.
\item The variation of permeability with $x$ in every case showed an upper bound for the cubic symmetry and a lower bound for the hexagonal packing. The values of $k$ against $x$ for the stochastic systems lay between these bounds, Fig.(\ref{k-l_cH})
\end{itemize}

 Our next step was to try and determine a single relationship between flow and the geometrical characteristics of the systems studied. Borrowing from the Kozeny-Carman equation \cite{Costa2006} and  permeability studies in \cite{Coussot2000}, and keeping in mind the Minkowski functionals, we proposed a relation between permeability and the geometrical parameters based on the behaviours obtained thus far:  
 \begin{equation}
 k = A_{c} \frac{\phi_{0}^{3}}{(1 - \phi_{0})^{2}} \frac{1}{{|\chi|}^{0.5}} \frac{1}{H}
 \label{form1}
 \end{equation}
where $A_{c}$ is the critical cross-section that cuts off particles with cross-sections greater than $A_{c}$ from percolating through the system; $\phi_{0}$ is the effective porosity, i.e., the porosity associated with the system spanning channels. The R.H.S. of Eq.(\ref{form1}) can be clubbed together as a combinatoric $F = A_{c} \frac{\phi_{0}^{3}}{(1 - \phi_{0})^{2}} \frac{1}{{|\chi|}^{0.5}} \frac{1}{H}$.  Following Eq.(\ref{form1}), we plotted the variation of the permeability $k$ with $F$ on a log-log scale as shown in Fig.(\ref{k-chi2_collapse}a). Though there was a good suggestion of the collapse of all points for all four scenarios studied onto a single straight line, there remained a non-negligible scatter. However, for all the ordered and disordered systems studied permeability followed a scaling relationship of the form

 \begin{equation}
 k = C {F }^{0.428}
 \label{fn_proposed}
\end{equation}

This indicated that irrespective of the disorder in pore space for 3-dimensional systems built of spherical grains, permeability followed a scaling behaviour with a combinatoric of the Minkowski functionals - $A_{c}$, $\phi_{0}$, $\chi$ and $H$, with a unique scaling exponent $0.428$.  The other notable part of our investigation is that the cubic and the hexagonal systems appear to provide the upper and lower bounds within which all values of $k$ reside for their corresponding combinatoric $F$. While a more rigorous proof is required for the bounds to be established, one can argue that given a fixed box size (here $L^{3}$ ), hexagonal close packing will certainly minimize the pore volume and, therefore, permeability. On the other hand, cubic packing with uniform-sized spheres shall provide the maximum void space and increase $k$. Any other distribution of particle size is expected to show permeability values between these limits.

To obtain an almost perfect collapse of all data points onto a single straight line, we followed a regression of Eq. (\ref{form1}) of the form
\begin{equation}
 k = \alpha (A_{c})^{a}\left(\frac{\phi_{0}^{3}}{(1-\phi_{0})^{2}}\right)^{b} ({|\chi|})^{c}H^{d}
 \label{regress}
\end{equation} %      
with $\alpha = 0.0081, a= -0.328; b = 1.061; c = 0.075$, $d = -1.602$. If we represent the RHS of Eq.(\ref{regress}) by $F^\prime$, Fig.(\ref{k-chi2_collapse}b) shows the variation of $k$ with $F^\prime$. It shows the power-law  scaling of the form

 \begin{equation}
 k = C^{\prime} {\left(F ^{\prime}\right)}^{0.998}
 \label{univ}
\end{equation}
 We propose that the permeability of a 3-dimensional porous structure, irrespective of its pore size and distribution, follows a power-law scaling with the sample geometrical measures.
Since we dealt with both ordered and disordered porous systems with various sizes and distributions of particles, the exponent $0.428$ may be universal, at least for systems constructed with spherical grains.

\section{Conclusions}

Fluid flow through porous systems is a very important phenomenon that affects many aspects of our daily life. It is almost intuitive to expect that the transport properties of fluids, like permeability and conductivity, should depend on the pore space geometry.  Hadwiger's theorem had already stated that any 3-dimensional system formed by a union of convex solids could be characterized by a linear combination of, at the most, four geometric invariants, the Minkowski functionals.  This provided us the impetus to try to correlate permeability with the Minkowski functionals. 

The quest began with the generation of two categories of disordered porous systems built with spherical particles, one having a log-normal size distribution and another with a normal distribution about a particular mean size. At least six different mean sizes were chosen for this construction. To understand the effect of disorder in sizes, we also worked with two categories of perfectly ordered porous systems but with different symmetry of grain arrangement: cubic and hexagonal. To mimic natural systems, the grains were allowed to fall under gravity and settle into a state of equilibrium using DEM. We solved flow equations for incompressible fluid that was injected from one end of the system spanning pore clusters under a fixed pressure gradient. The steady-state velocity and pressure field were solved at every point of the transport system, and permeability was computed using Darcy's law.

Four geometric characteristics for every structure constructed were determined, these being (i) Critical cross-sectional area, (ii) Effective porosity, (iii) Integral mean curvature, and (iv) Euler Characteristic. With each of these four geometric characteristics, permeability showed power-law scaling. Though the effective porosity $\phi_{0}$ did not follow a power-law, we observed that the `cloud' of data points for all the systems considered followed a scaling behaviour with permeability. For each of the Minkowski Functionals explored, the exponent of the power-law was approximately identical for both, the disordered and deterministic systems.

Based on our progressive findings, we constructed a combinatorial function $F$ of the geometric characteristics inspired by Hadwiger's theorem. The unique finding of this work was that the permeability of the porous system showed a single scaling relation with the combinatoric $F$ of the four Minkowski functionals for all the systems studied, irrespective of their ordered or disordered pore geometry. The exponent of the scaling relation is $0.428$, which is universal for 3-dimensional porous systems constructed of spherical grains. To the best of our knowledge, this is the first time that a single relationship combining the fluid flow with all the Minkowski functionals has been proposed with a unique exponent for 3-dimensional disordered porous systems. A collapse of all the data points of all the four systems studied was achieved using a regression of the combinatoric $F$. 

Another interesting finding of the work is that the cubic and hexagonal systems formed the upper and lower bound of all the scaling relations between $k$ and each of the geometric measures studied. The data for the stochastic systems lay between these bounds.  
 
Our original goal of correlating permeability to system characteristics via four Minkowski functionals having a single value of exponent irrespective of size or distribution of particles suggests that the exponent is universal.
It remains to be seen whether the exponent $0.428$ is independent of particle shape. Other transport properties like electrical conductivity need to be explored along the same lines. We hope to be able to report our findings in this direction in the near future.
  
\begin{acknowledgments}
R.A.I.Haque is grateful to UGC for funding (UGC-Ref No. 1435/(CSIR-UGC NET JUNE 2017).
\end{acknowledgments}

\vskip 2cm

%\section*{References}
\nocite{*}
\bibliographystyle{elsarticle-num}
\bibliography{ref_perm}

\end{document}